\documentstyle[prl,aps,psfig]{revtex}
\newcommand{\be}{\begin{equation}}
\newcommand{\ee}{\end{equation}}
\newcommand{\bea}{\begin{eqnarray}}
\newcommand{\eea}{\end{eqnarray}}
\begin{document}

 \title{Exact Phase Diagram of a model with Aggregation and Chipping}
 \author{R. Rajesh$^1$ and Satya N. Majumdar$^{1,2}$}
 \address{ {\small 1. Department of Theoretical Physics, Tata Institute of
Fundamental Research, Homi Bhabha Road, Mumbai 400005, India.}\\
 {\small 2. Laboratoire de Physique Quantique (UMR C5626 du CNRS),
Universit\'e Paul Sabatier, 31062 Toulouse Cedex, France.}}
 \date{\today}
 \maketitle
 \widetext
 %%%%%%%%%%%%%%%%%%%%%%%%%%%%%%%%%%%%%%%%%%%%%%%%%%%%%%%%%%%%%%%%%%%%%%%%    

\begin{abstract} 
 We revisit a simple lattice model of aggregation in which masses diffuse
and coalesce upon contact with rate $1$ and every nonzero mass chips off a
single unit of mass to a randomly chosen neighbour with rate $w$. The
dynamics conserves the average mass density $\rho$ and in the stationary
state the system undergoes a nonequilibrium phase transition in the
$(\rho-w)$ plane across a critical line $\rho_c(w)$. In this paper, we
show analytically that in arbitrary spatial dimensions, $\rho_c(w) =
\sqrt{w+1}-1$ exactly and hence, remarkably, independent of dimension. We
also provide direct and indirect numerical evidence that strongly suggest
that the mean field asymptotic answer for the single site mass
distribution function and the associated critical exponents are
super-universal, i.e., independent of dimension. 

\vskip 5mm
\noindent
PACS numbers: 64.60.-i, 05.70.Ln   
\end{abstract}
%%%%%%%%%%%%%%%%%%%%%%%%%%%%%%%%%%%%%%%%%%%%%%%%%%%%%%%%%%%%%%%%%%%%%%%%%%
\section{Introduction}

Nonequilibrium phase transitions \cite{hinrichsen} occur in various
systems including heterogeneous catalysis \cite{ziff}, chemical reaction
models \cite{schogl}, polynuclear growth models \cite{kertesz},
monomer-dimer models \cite{kim}, models of fungal growth \cite{lopez},
non-equilibrium kinetic Ising models \cite{odor}, and branching
annihilating random walks \cite{cardy}.  A common feature in all the above
is that the transition is from a state that has no activity to one that
has continued activity. Such transitions are well characterized by the
critical exponents of Directed Percolation, Parity Conserving, or the DP2
universality classes. There are other nonequilibrium models that undergo
phase transitions which do not belong to the above universality classes. 
These include boundary driven phase transitions \cite{derrida} and models
whose steady states undergo a `de-pinning' or `unbinding'
transitions\cite{hin2,hm,mkb3,gm}. Recently a simple lattice model where
masses diffuse, aggregate on contact, and also chips off a single unit of
mass was studied\cite{MKB1,MKB2}.  This `chipping' model (CM)  exhibits a
nonequilibrium phase transition\cite{MKB1,MKB2} in its steady state from a
phase in which an {\it infinite} aggregate is present to one that has
none. This nonequilibrium phase transition is in a completely different
universality class compared to the other models studied in the literature
and mentioned above. The mathematical mechanism giving rise to the
formation of the infinite aggregate at the onset of phase transition was
found to be very similar to that of the equilibrium Bose-Einstein
condensation in an ideal Bose gas. The difference is that in CM, the
infinite aggregate or the condensate forms in real space as opposed to the
Bose gas where the condensation takes place in the momentum space.
Besides, the phase transition in the CM occurs even in $1$-dimension as
opposed to the Bose gas where the condensation occurs in $2$ and higher
dimensions. A slightly different off-lattice version of the CM was studied
earlier within rate equation approach in the context of aggregation in dry
environments\cite{kr}. A directed version of the CM where masses move
asymmetrically only along one direction was studied in Ref.\cite{MKB2}
and its critical properties were found to belong to a different university
class from that of the undirected CM. This directed CM also appeared
recently in the context of a traffic model with passing\cite{IK}. 
  
The undirected CM is defined on a $d$-dimensional hyper cubic lattice with
periodic boundary conditions. Starting from a random distribution of
nonnegative integer masses at each lattice site, the system evolves in
continuous time via the following microscopic processes. In an
infinitesimal time interval $\Delta t$, (i)  with probability $\Delta t$,
the mass at each site hops to one of its neighboring sites, chosen at
random, (ii) with probability $w\Delta t$, a unit mass is chipped off from
an already existing mass at each site and added to one of the neighboring
sites, again chosen at random, and (iii) with probability $(1-(1+w)\Delta
t)$, each mass stays at its original site. Following the steps (i)-(iii),
the masses at any given site add up.  Since this system is closed to the
environment (periodic boundary condition), the total mass is conserved by
the dynamics. Thus, there are two parameters in the problem, the average
mass per site $\rho$, and the ratio of the chipping rate to the rate of
hopping as a whole, $w$.  In the long time limit, the system evolves into
a time independent steady state.  The steady state single site mass
distribution function $P(m)$, i.e., the probability that a site has mass
$m$ when $t\to \infty$, was shown to undergo a phase transition in the
$\rho$-$w$ plane\cite{MKB1,MKB2}. There is a critical line $\rho_c(w)$ in
the $\rho$-$w$ plane that separates two types of asymptotic behaviors of
$P(m)$.  For fixed $w$, as $\rho$ is varied across the critical value
$\rho_c(w)$, the large $m$ behaviour of $P(m)$ was found to be,
 \be
 P(m)\sim \cases
          {e^{-m/m^{*}} &$\rho<\rho_c(w)$,\cr
          m^{-\tau} &$\rho=\rho_c(w)$,\cr
          m^{-\tau}+ {\rm infinite\,\,\, aggregate} &$\rho>\rho_c(w)$.\cr}
 \label{pm}
 \ee 
 Thus, the tail of the mass distribution changes from having an
exponential decay to an algebraic decay as $\rho$ approaches $\rho_c$ from
below. As one increases $\rho$ beyond $\rho_c$, this asymptotic algebraic
part of the critical distribution remains unchanged but in addition an
infinite aggregate forms. This means that all the additional mass
$(\rho-\rho_c)V$ (where $V$ is the volume of the system) condenses onto a
single site and does not disturb the background critical distribution. 
This is analogous, in spirit, to the condensation of a macroscopic number
of bosons onto the single $k=0$ mode in an ideal Bose gas as the
temperature goes below a certain critical value. This infinite tower,
i.e., the single site carrying macroscopically large (proportional to
volume) mass is mobile.  Mathematically this means that for
$\rho>\rho_c(w)$, $P(m)\sim m^{-\tau} + {1\over {V}}\delta(m-\alpha V)$
where $\alpha=\rho-\rho_c$. This last delta-function peak contributes only
to order $1/V$ in the integral $\int P(m)dm$ indicating that the aggregate
is contained in thermodynamically few sites (indeed a single site) but
contributes a finite amount $\alpha$ in the integral $\int mP(m)dm$ even
in the thermodynamic limit $V\to \infty$. 

These results were found both analytically within a mean field
approximation which involved ignoring all correlations between masses, and
numerically in one dimension. Within the mean field approximation, the
locus of the critical line was found\cite{MKB1} to be, $\rho_c(w)=\sqrt
{w+1}-1$ and the exponent $\tau=5/2$\cite{MKB1,kr}. In one dimension, the
numerically obtained critical line\cite{MKB1} was found to be close to the
mean field phase curve. Even the exponent $\tau$, determined from a simple
linear fit on the log-log plot of data from relatively smaller size
lattices, was found to be $\tau\approx 2.33$, rather close to the mean
field exponent $2.5$. This raises the question whether or not the mean
field answers for asymptotic behaviors of $P(m)$ are exact even in one
dimension. On the other hand, one can show explicitly (see section II) 
that there exist nonzero correlations between masses in any finite
dimension, even in the thermodynamic limit. Thus, one is confronted with a
puzzle. 

The purpose of this paper is to shed new light on this puzzling issue. We
first prove analytically a remarkable result that the mean field phase
boundary, $\rho_c(w)=\sqrt {w+1}-1$, is indeed {\it exact} and independent
of the spatial dimension $d$. This, of course, still does not prove, but
hints, that the exponent $\tau$ may also be independent of $d$.  However,
we provide rather unambiguous numerical evidence, in conjunction with
several direct and indirect checks, which suggest that even the exponent
$\tau=5/2$ is super-universal and independent of $d$. Thus, our results
seem to suggest strongly that even though there are nonzero correlations
between masses in finite dimensions, these correlations do not affect the
asymptotic behavior of the single site steady state mass distribution
$P(m)$. However, the possibility remains that other higher order
correlation functions, such as the joint distribution of two masses
$P(m_1,m_2)$ will depend on the spatial dimension $d$. 
  
The paper is organized as follows. In section II, we show analytically
that in arbitrary dimensions $d$, the locus of the phase boundary is
independent of $d$ and is given by the mean field expression,
$\rho_c(w)=\sqrt {w+1}-1$. In section-III, we re derive analytically the
mean field expression for the mass distribution $P(m)$ by a method
different from that used in \cite{MKB1} and compare these analytical
expression with the numerical results obtained in one and two dimensions. 
In section IV, we do a finite size scaling analysis that provides
additional evidence that the exponent $\tau$ is super-universal. Finally,
we conclude with a summary and discussion in section V. 

\section {Exact Phase Diagram in Arbitrary Dimensions}

As mentioned in the introduction, the steady state of the CM undergoes a
phase transition in its parameter space ($\rho$-$w$ plane) across a
critical line $\rho_c(w)$ in all dimensions. In this section, we compute
$\rho_c(w)$ exactly in arbitrary dimensions by analyzing the two point
equal time mass-mass correlation function, $C({\bf x},t)= \langle m({\bf
x},t)m({\bf 0},t) \rangle $. Let $\eta(\bf x,\bf x',t)$ denote the mass
transferred from a site $\bf x$ to a neighboring site $\bf x'$ in the time
interval between $t$ and $t+\Delta t$.  Clearly, $\eta(\bf x,\bf x',t)$ is
a random variable that takes the following values,
 \be
  \eta(\bf x,\bf x',t)= \cases
      {m(\bf x),\,\,\,\, {\rm with\,\,\, prob.} &${1\over {2d}}\Delta t$,\cr
       1-\delta_{m({\bf x}),0} , \,\,\,\,{\rm with \,\,\, prob.} 
	   &${1\over {2d}}w\Delta t$,\cr
       0 ,\,\,\,\, {\rm with \,\,\, prob.} &$\left(1- {{1+w}\over {2d}}
	   \Delta t\right )$,  \cr}
 \label{trans}
 \ee      
 where $2d$ is the number of neighbors of any given site. The Kronecker
delta function in the second line of Eq. (\ref{trans})  indicates that a
chipping of a unit mass can take place provided the mass $m(\bf x)$ is a
positive integer bigger than $0$.  Then, the evolution of mass at site
${\bf x}$ can be written as,
 \be
 m({\bf x},t+\Delta t)=m({\bf x},t)-\sum_{\bf x'}
 \eta ({\bf x},{\bf x'},t) +
 \sum_{\bf x'} \eta ({\bf x'},\bf x,t),
 \label{one}
 \ee 
 where the sum is over the neighbors ${\bf x'}$ of the site ${\bf x}$. 
The second term on the right hand side of Eq. (\ref{one}) describes the
outflow of mass from ${\bf x}$ while the third term accounts for the
inflow of mass into ${\bf x}$ from neighboring sites. It is quite
straightforward to write down the $2$ point correlator for $\eta$ to order
$\Delta t$.  Suppressing the explicit $t$ dependence in $\eta$ we find,
 \be
 \langle \eta ({\bf x}_1,{\bf {x'}}_1) \eta ({\bf x}_2, {\bf x'}_2)\rangle = 
 {1\over {2d}}\left( {m}^2( {\bf x}_1)+w(1-\delta_{m({\bf x}_1),0})\right)
 \delta_{{\bf x}_1,{\bf x}_2}\delta_{{\bf x'}_1,{\bf x'}_2} \Delta t .
 \label{two}
 \ee
 Using Eqs. (\ref{one}) and (\ref{two}), the evolution equations for the
two-point equal time correlation function, $C({\bf x},t)= \langle m({\bf
x},t)  m({\bf 0},t) \rangle $ can be written down. Multiplying Eq. 
(\ref{one}) by $m({\bf 0},t+dt)$ and taking average on both sides, and
putting all time derivatives to zero in the steady state, we get,
 \bea
 -C({\bf x})+ {1\over {2d}}\sum_{\bf x'} C(\bf x') 
 &=&
 w \left(-D({\bf x})+{1\over {2d}}\sum_{\bf x'} D(\bf x')  \right) 
\nonumber \\
 &-&\left({C({\bf 0})+ w s} \right)\left ( \delta_{{\bf
 x},0}-{1\over {2d}}\sum_{\bf x_0}\delta_{ {\bf x},{\bf x_0}}
 \right) ,
 \label{three}
 \eea
 where ${\bf x_0}$ denotes the neighbors of the site ${\bf 0}$. Also,
$D({\bf x},t)= \langle m({\bf x},t) \delta_{m({\bf 0},t),0}\rangle$, and
$s=1-\langle \delta_{m({\bf 0}),0}\rangle$, is the probability of a site
having non zero mass. Thus, in the CM, the two point correlation functions
do not form a closed set of equations making it difficult to solve for
$C(\bf x,t)$ exactly. This is unlike many other models where two point
correlations do form a closed set of equations and hence are solvable.  A
few examples include the $1$-d Glauber model\cite{Glauber}, asymmetric
random average process\cite{krug,RM}, the Takayasu model of
aggregation\cite{RM1} and the $q$ model of force fluctuations in bead
packs \cite{RM1,LMY}. 

Remarkably, however, Eq. (\ref{three}) allows for the solution,
 \be
 C({\bf x})  = w [ D({\bf x})-s]\quad, \qquad {\bf x}\neq {\bf 0}. 
 \label{four}
 \ee
 This solution is also the unique solution. To see this, observe that the
homogeneous part of Eq. (\ref{three}) is the Laplace's equation
$\nabla^2 (C({\bf x})-w D({\bf x}))=0$, with the boundary condition
that $(C({\bf x})-w D({\bf x}))$ is a constant as $\mid {\bf x}\mid
\rightarrow \infty$. Since the solution Eq. (\ref{four}) satisfies the
inhomogeneous part too, as well as the boundary conditions, it is the unique
solution.

Note that the above solution Eq. (\ref{four}) is also valid on a finite 
lattice.
Summing Eq. 
(\ref{four}) over all $\bf x\neq 0$ and using the fact that the conserved
total mass is given by, $\sum_{\bf x}m({\bf x})=\rho V$ (where $V=L^d$ is
the volume of the system), we get the following exact equation,
 \be
 \rho^2-{ {\langle m^2\rangle}\over {V}}=w \rho(1-s)-w s. 
 \label{eleven}
 \ee
 This equation is reminiscent of Bose-Einstein condensation in ideal Bose
gas. In the low density phase, we expect that the system reaches a
stationary state in which $\langle m^2\rangle$ is a finite number of order
$O(1)$.  Therefore, in the thermodynamic limit $V\to \infty$, the second
term on the left hand side of Eq. (\ref{eleven})  drops out and we get,
 \be
 \rho^2=w \rho(1-s)-w s. 
 \label{five}
 \ee
 Note that the Eq. (\ref{five}) could have been obtained from Eq. 
(\ref{four}) if we had assumed that the two point correlation functions
decouple, i.e., $C({\bf x})=\langle m({\bf 0})m({\bf x})\rangle =\langle
m({\bf 0})\rangle \langle m({\bf x})\rangle={\rho}^2$ and $D({\bf x})=
\langle m({\bf x})\delta_{m({\bf 0}),0}\rangle=\rho(1-s)$. However, there
is apriori no reason for the two point correlations to decouple.  Our
derivation of Eq. (\ref{five}) {\it does not} rely on this decoupling. 

From Eq. (\ref{five}) we get,
 \be
 s=\frac{w \rho-\rho^2}{w(\rho+1)}. 
 \label{six}
 \ee
 According to Eq. (\ref{six}), as one increases the density $\rho$ keeping
$w$ fixed, the occupation probability $s$ first increases with $\rho$,
attains a maximum at $\rho=\sqrt {w+1} -1$ (obtained by setting
$\frac{ds}{d\rho}=0$ in Eq.  (\ref{six})) and then starts decreasing with
increasing $\rho$. However, it is clear that $s$, the probability that a
site has nonzero mass, must be a monotonically non decreasing function of
$\rho$.  Hence we conclude that Eq. (\ref{six}) is valid as long as
$\rho\leq \rho_c=\sqrt{w+1}-1$. For $\rho>\rho_c$, the basic assumption
$\langle m^2\rangle /V \to 0$ as $V\to \infty$ breaks down and Eq. 
(\ref{six}) ceases to be valid. 

Thus, the critical density is given by,
 \be
 \rho_c(w)=\sqrt{1+w}-1 ,
 \label{eight}
 \ee
 and remarkably, it is independent of $d$ and not surprisingly, therefore,
coincides with the mean field expression\cite{MKB1,MKB2}.  For $\rho\leq
\rho_c$, $s$ is given by Eq. (\ref {six}). As $\rho$ increases from $0$ to
$\rho_c(w)$ (for fixed $w$), $s$ increases monotonically according to Eq. 
(\ref{six}) up to the value $s_c$ given by,
 \be
 s_c=\frac{\sqrt{1+w}-1}{\sqrt{1+w}+1} . 
 \label{nine}
 \ee
 For $\rho>\rho_c(w)$, $s$ does not increase any further and sticks to its
value $s_c$. Putting $s=s_c$ in Eq. (\ref{eleven}) and using the
expression of $s_c$ from Eq. (\ref{nine}), we get for $\rho>\rho_c(w)$,
 \be
 \lim_{V\rightarrow \infty} \langle \frac{m^2}{V}\rangle =(\rho-\rho_c)^2
.
 \label{twelve}
 \ee
 Thus, for $\rho>\rho_c$, $\langle m^2\rangle$ becomes macroscopic, i.e,
proportional to volume.  Since $s$, the fraction of occupied sites, does
not increase anymore for $\rho>\rho_c$, this indicates that all the extra
mass $(\rho-\rho_c)V$ condenses onto a thermodynamically negligible number
of sites (indeed, a single site only)
with density $\sim {1\over {V}}$, leading to the macroscopic
behavior of $\langle m^2\rangle \sim {1\over {V}}[(\rho-\rho_c)V]^2\sim
(\rho-\rho_c)^2 V$.  This is similar in spirit, though not in details, to
the Bose-Einstein condensation where below a certain temperature, the
number of particles in the $k=0$ mode also become macroscopic. 

Let us conclude this section by stressing on an important point. We note
that the mean field solution (assuming decoupling) for the stationary two
point correlation function, $C({\bf x})  =\rho^2$ for ${\bf x}\neq {\bf
0}$ and $D({\bf x})=\rho (1-s)$ for ${\bf x}\neq {\bf 0}$ with $s$
satisfying Eq. (\ref{eleven}), is indeed an exact solution of Eqs. 
(\ref{three})  and (\ref{four}). However, this need not be the
only stationary solution. Besides, even if the mean field answer for the
two point stationary correlation function is the correct one, it still
does not prove that the mean field theory is exact. For example, one can
show\cite{RM2} that indeed $3$ and higher point correlation functions do
not decouple. In any case, the main result of this section, namely the
derivation of the exact phase boundary, does not rely on whether the
correlation functions decouple or not. 
 
\section{Comparison with Mean Field Theory}

In the previous section, we proved that the mean field phase diagram is
exact in any dimension. This, of course, does not prove but suggests that,
perhaps, even the mean field expression for the distribution $P(m)$ may
also be asymptotically exact in all dimensions.  In this section, we try
to provide evidence in favor of this hypothesis. For this we first re
derive the mean field expression for $P(m)$ and compare it with the
numerical results obtained in $1$ and $2$ dimensions. 

In Ref. \cite{MKB1}, the steady state single site mass distribution
function $P(m)$ was computed analytically by assuming that the joint
distribution $P(m_1,m_2)$, the probability that two consecutive sites have
masses $m_1$ and $m_2$ respectively, factorises, i.e.,
$P(m_1,m_2)=P(m_1)P(m_2)$. With this assumption, $P(m)$ was shown to
satisfy a closed set of equations which were then solved via generating
function method yielding results mentioned in Eq. (\ref{pm})  with
$\tau=5/2$ and $\rho_c(w)=\sqrt{w+1}-1$. In this section, we first re
derive the mean field results by a different method that requires lesser
restrictions than the product measure used in Ref. \cite{MKB1}. 

Here we use a technique used before for solving the mass distribution
function in other models of aggregation\cite{takayasu} as well as the $w=
0$ limit of the CM\cite{MH}.  We consider the CM on a $1$-d lattice. Let
$P(m_1,m_2,\ldots , m_n)$ denote the joint probability that $n$
consecutive sites on the lattice have masses $m_1$, $m_2$, $\ldots$, $m_n$
respectively in the stationary state.  We define two generating functions,
 \bea
 Z_n&=& \sum_{m_1=0}^{\infty} \ldots \sum_{m_n=0}^{\infty}
 x^{m_1+\ldots+m_n} P(m_1,\ldots,m_n), \\
 Y_n&=&\sum_{m_1=0}^{\infty} \ldots \sum_{m_n=0}^{\infty}
 x^{m_1+\ldots+m_n} P(m_1,\ldots,m_n,0). 
 \label{mf1}
 \eea
 Here $Z_n=\langle x^{m_1+\ldots+m_n} \rangle$ is an unconditional average
but $Y_n={\langle x^{m_1+\ldots+m_n}\rangle}_0$ is a conditional average
where the $(n+1)$-th site is conditioned to have $0$ mass. Using the
dynamics of $m_i$'s (as given by Eq. (\ref{one})) and following steps
similar to those used in Refs. \cite{takayasu,MH}, one can write down the
evolution equations for $Z_n$'s.  In the steady state, when all time
derivatives go to zero, we get, after some algebra,
 \be
 Z_{n+1}-2 Z_n+Z_{n-1}+w \left( \frac{(1-x)^2}{x}
Z_n-\frac{1-x}{x}Y_{n-1}+
 (1-x) Y_n) \right)=0,
 \label{mf2}
 \ee
 with the boundary conditions, $Z_0=1$ and $Y_0=P(0)=1-s$, the probability
of having no mass at any given site. 

If we now make the assumption that $P(m_1+\ldots+m_n=m,0)=
P(m_1+\ldots+m_n=m)P(0)$, i.e., $Y_n=P(0) Z_n$, then we get equations that
contain only the $Z_n$'s. As mentioned before, this assumption is less
strict than the product measure approximation as it requires the
factorization of only a special conditional probability where a site at
the beginning of a string is empty. In order to determine $P(m)$, we need
to compute $Z_1(x)$ which, by definition in Eq. (\ref{mf1}) is the
generating function for $P(m)$'s,
 \be
 Z_1(x)=\sum_{m=0}^{\infty}x^m P(m). 
 \label{mfZ1}
 \ee
 Since $Z_1$ depends on other $Z_n$'s, we need to solve the full Eq. 
(\ref{mf2}) for all $n$.  Eq. (\ref{mf2}) can be solved by standard
generating function method.  Let $G(x,y)=\sum_{n=1}^{\infty} Z_n(x) y^n$. 
Multiplying Eq. (\ref{mf2}) by $y^n$ and summing over $n$, we get, after
straightforward algebra,
 \be
 G(x,y)=\frac{y \left[w P(0) y (1-x)-x y+x Z_1(x) \right]}{x(1-y)^2+w y
(1-x)^2+w P(0) y
 (1-x) (x-y)}. 
 \label{mf3}
 \ee
 For a fixed $x$, when considered as a function of $y$ only, $G(x,y)$ has
two poles,
 \be
 y_{1,2}= \frac{-2 x+w (1-x) (1-s x)\pm (1-x)\sqrt{w} \sqrt{w (1-s x)^2-4
s
 x}}{2\left((1-s)w (1-x)-x\right)}.
 \label{mf4}
 \ee
 
For a fixed $w$ and $s$, $\mid y_2\mid<1$ for small values of $x$. This
implies that $Z_n \sim \mid y_2\mid^{-n}$ for large $n$. However we cannot
have a diverging probability for large $n$. Hence this pole must be
canceled off by the numerator of $G$. This `pole canceling' mechanism was
also useful in deriving exact results in other recently studied
aggregation models\cite{RM,RM1}.  Demanding $G(x,y_2)=0$, we get the
following expression for $Z_1(x)$,
 \be
 Z_1(x)=\frac{2x-w (1-x) (1-s x)+(1-x)\sqrt{w}\sqrt{w (1-s x)^2-4 s x}}{2
x}.
 \label{mf5}
 \ee
 The coefficient of $x^m$ on the right hand side of Eq. (\ref{mf5}) will
then give the desired distribution $P(m)$. 

The expression for the generating function $Z_1(x)=\sum_{m=0}^{\infty}
x^mP(m)$ in Eq. (\ref{mf5}) is identical to the one derived in \cite{MKB1}
using the approximation of product measure. Thus, the two methods, though
different in details, yield the same $P(m)$ for {\it any} $m$ and not just
for large $m$. Thus, this result for mean field $P(m)$ seems to be
extremely robust and does not depend on the details of how the mean field
assumption is incorporated. 

The asymptotic properties of $P(m)$ for large $m$ can be derived by
analyzing the behaviour of $Z_1(x)$ near $x=1$\cite{MKB1} and one recovers
the results in Eq. (\ref{pm}) with $\tau=5/2$ and
$\rho_c(w)=\sqrt{w+1}-1$. However, by expanding the expression for
$Z_1(x)$ in Eq. (\ref{mf5}) in powers of $x$ using Mathematica, we have
determined $P(m)$ for all $m$. In the aggregate phase ($\rho>\rho_c$), we
set $s=s_c$ in Eq. (\ref{mf5}) and calculate the distribution $P(m)$ by
expanding in powers of $x$. In Fig. (1), we compare this analytical mean
field answer for all $m$ with the numerical results obtained in $1$ and
$2$ dimensions in the aggregate phase ($\rho> \rho_c$). Note that Fig. (1) 
shows only the power law part of the spectrum. The numerical data is for
$V=900$ in $1$-dimension and for a $30X30$ lattice in $2$-dimensions.  The
two curves for $1$ and $2$-d are almost indistinguishable from each other.
While the numerical data matches excellently with the mean field answer
for small $m$, it has a small deviation for larger masses. This deviation
at large $m$ is due to finite size effects.  To confirm this, we also did
simulations for larger lattice sizes up to $V=2000$ in $1$-dimension. For
a fixed mass, we confirmed that the deviation from the mean field
decreases with increasing $V$ (see the panel inside Fig. (1)).  We also
compared the mass distribution in the exponential phase ($\rho<\rho_c$)
with the mean field prediction. Again excellent agreement is seen (see
Fig. (2)). These results thus provide strong evidence that the mean field
expression for $P(m)$ is exact in all dimensions and therefore the
exponent $\tau=5/2$ is also super-universal, i.e., independent of $d$. 

\section{Finite Size Scaling and Indirect Numerical Checks}

In this section, we provide further indirect numerical checks which again
strongly suggests that the mean field answer for $P(m)$ is indeed exact. 
We start by making a reasonable finite size scaling ansatz for $P(m,V)$
(where $V$ is the volume of the system) in the aggregate phase
$\rho>\rho_c$ for large $m$,
 \be
 P(m,V)\approx \frac{1}{m^{\tau}}f\left(\frac{m}{V^{\phi}} \right) 
 +\frac{1}{V} \delta \left(m-(\rho-\rho_c) V\right),
 \label{ten}
 \ee
 where the exponent $\phi$ is a crossover exponent and the delta function
peaked at $(\rho-\rho_c) V$ indicates the aggregate containing a
macroscopic amount of mass. The power-law part of the mass distribution
gets cut off at $\sim V^{\phi}$ for finite $V$. Since the largest mass
inside the power law part is much less compared to that in the aggregate,
we get the inequality, $\phi<1$. 

Also, since the mass density, $\int mP(m)dm=\rho$ is finite. Substituting
Eq. (\ref{ten}) in this integral, it is evident that for large $V$, the
integral will converge provided $\tau>2$. Also, from Eq. (\ref{ten}), one
finds that the second moment, ${ {\langle m^2\rangle}\over {V}} \approx
(\rho-\rho_c)^2 + O(V^{\phi(3-\tau)-1})$. The first term is from the
aggregate and the second term is from the power-law part. This is
consistent with the exact result Eq. (\ref{twelve}) provided,
$\phi(3-\tau)<1$.  This provides an upper bound for $\tau$, i.e.,
$\tau<3$.  Thus, we have the exact bounds for $\tau$, $2<\tau<3$. The mean
field results as well as numerical simulations in $1$ and $2$-d are
consistent with these bounds. 

Next we derive a scaling relation between $\tau$ and $\phi$. This is
obtained by demanding that $P(m,V)$ is normalized, $\int P(m,V)dm =1$. 
Substituting Eq. (\ref{ten}) in this integral, we get,
 \be
 \int_{m_l}^{\infty}{1\over {m^{\tau}}}f({m\over {V^{\phi}} })dm
=c-{1\over {V}},
 \label{fs}
 \ee
 where $c$ is a constant of $O(1)$, and $m_l\sim O(1)$ , is the mass
beyond which the scaling starts holding. Differentiating Eq. (\ref{fs}) 
twice with respect to $V$ and making a change of variable, we arrive at
the relation,
 \be
 \int_{m_lV^{-\phi}}^{\infty}f''(y)y^{2-\tau}dy =
-(1-\phi)V^{\phi{\tau-1}-1},
 \label{fs0}
 \ee
 where $f''(y)={ {d^2f}\over {d^2y}}$. The reason for twice
differentiation is as follows. Since $2<\tau<3$, one can replace the lower
limit of the integral in Eq. (\ref{fs0}) by $0$ in the $V\to \infty$ limit
and there is no divergence from the lower cut off. Once the lower cut off
is replaced by $0$, the integral on the left hand side of Eq. (\ref{fs0}) 
is of $O(1)$. Comparing this with the right hand side of Eq. (\ref{fs0}),
we immediately arrive at the scaling relation,
 \be
 \phi (\tau-1)=1. 
 \label{fs1}
 \ee
 If $\tau=5/2$, Eq. (\ref{fs1}) would indicate $\phi=2/3$. 

We found that the cleanest way to measure the exponent $\phi$ is via the
following indirect finite size method.  On a finite lattice, the critical
value of the occupation probability of a site, $s_c(V)$ will differ from
the $V=\infty$ value, $s_c(\infty)$. One can assume a reasonable finite
size correction of the form,
 \be
 s_c(V)=s_c^{\infty}-\frac{a}{V^{\theta}},
 \label{thirteen}
 \ee
 where $a$ is a constant and $\theta$ is a new exponent. Using this ansatz
along with the expression of $\langle m^2\rangle$ obtained from Eq. 
(\ref{ten}) in Eq. (\ref{eleven}), we find $V^{\phi(3-\tau)-1} \sim
V^{-\theta}$, implying another scaling relation,
 \be
 \theta=1-\phi(3-\tau). 
 \label{fs2}
 \ee
 If $\tau=5/2$ and $\phi=2/3$, we get from Eq. (\ref{fs2}), $\theta=2/3$. 
In Fig. (3), we plot, in one dimension, $s_c(V)$ for various values of $V$
and indeed find that $\theta=2/3$. This provides an indirect numerical
check on these scaling assumptions as well as suggests, once again, that
$\tau=5/2$ is a super-universal exponent. 

Another indirect check can be done by mapping the one dimensional CM to a
fluctuating interface model\cite{MKB2}.  We outline the procedure in
brief. The first step is to map the CM on to a system of hard core
particles moving on a ring. For this, interpret $m_i$, the mass at site
$i$, to be the gap between the $(i-1)^{th}$ and $i^{th}$ particle.  Then
the chipping move corresponds to a particle jumping forward one step
provided the target site is empty, while the aggregation move corresponds
to a particle making a long range jump to the site adjacent to the
particle nearest to it. There is a standard procedure to map a lattice gas
configuration to an interface configuration\cite{spohn}.  Let $n_j=1 (-1)$
if a particle is present (absent) at site $i$. Then $h_i=\sum_{j=1}^{i}
n_j$. While the chipping move corresponds to local moves of the interface,
the aggregation move corresponds to nonlocal moves of the interface. The
width of this interface was monitored numerically\cite{MKB2} as a function
of time $t$. On the critical line $\rho_c(w)$ of the CM, the width was
found to have the scaling form\cite{MKB2},
 \be
 w(V,t)\approx V^{\chi} f\left( \frac{t}{V^z}\right),
 \label{fs3}
 \ee
 with $\chi \approx 0.67$ and $z\approx 2.0$. 

We show here that the roughness exponent $\chi$ can be related to the
exponents $\tau$ and $\phi$ of the CM. We first map the CM in $1$-d to an
interface model with the height field $h_i=\sum_{j=1}^{i} m_j$. Then
$\langle h_i \rangle =i \rho$ and $\langle h_i^2 \rangle \approx i\langle
m^2 \rangle + i (i-1) \rho^2$. If we approximate $\langle m_i m_j \rangle$
by $\rho^2$ (assuming that the mean field $P(m)$ is exact), the width is
then simply $w^2=1/V \sum_{i=1}^V (h_i -\langle h_i \rangle)^2 \sim
V^{1+\phi (3-\tau)}=V^{2\chi}$. This implies that provided $\langle m_i
m_j \rangle\approx {\rho}^2$, one gets the scaling relation,
 \be
 \chi={1\over {2}}[1+\phi (3-\tau)]. 
\label{fs4}
 \ee
 If $\tau=5/2$ and $\phi=2/3$, we would get from Eq. (\ref{fs4}),
$\chi=2/3$, in excellent agreement with the numerical value $\chi \approx
0.67$\cite{MKB2}.  This is further evidence for $\tau=5/2$ and $\phi=2/3$. 

\section{Summary and Outlook}

In this paper, we have studied a simple stochastic lattice model where
masses diffuse as a whole and coalesce upon contact with rate $1$ and
every nonzero mass chips off a single unit of mass to a neighbour with
rate $w$. The mass density $\rho$ is conserved by the dynamics.  This
model undergoes a nonequilibrium phase transition in the $(\rho-w)$ plane
across a critical line $\rho_c(w)$. We have shown analytically that
$\rho_c(w)=\sqrt{w+1}-1$ in all dimensions. We also provided both direct
and indirect numerical evidence that strongly suggest that the mean field
answer for the single site mass distribution function $P(m)$ might be
exact in all dimensions and that the exponent $\tau=5/2$ is
super-universal. 

However we would like to stress one important point. Even though the
single site distribution $P(m)$ may be given exactly by the mean field
answer, that does not prove that mean field theory or product measure is
the exact stationary state in all dimensions. In this sense, the CM is
different from the recently studied $q$-model of force
fluctuations\cite{coppersmith} where the product measure is exact in the
stationary state. More precisely, in the $q$-model, the evolution equation
for single site mass distribution $P(m)$ involves the $2$-point
distribution $P(m_1,m_2)$. Similarly the equation for $2$-point involves
the $3$-point function $P(m_1,m_2,m_3)$ and so on. However, if one assumes
the product measure, i.e., $P(m_1,m_2,\ldots m_n)=P(m_1)P(m_2)\ldots
P(m_n)$ for all $n$, then it was shown\cite{coppersmith} that this ansatz
satisfies all the equations of the hierarchy. In the CM, in a similar way,
one can write down the full hierarchy of equations satisfied by the
$n$-point distribution functions. However, unlike the $q$-model, the
product measure ansatz does not satisfy all the equations of this
hierarchy\cite{RM2}. Nevertheless, the expression of $P(m)$ obtained from
the first equation of this hierarchy (the equation that involves only
$P(m)$ and $P(m_1,m_2)$) seems to be extremely close to the numerical
answer. This suggests that in the stationary state of CM, correlations
between masses appear only in the $3$ or higher order correlations but
seem to be absent at the $2$-point level.  This remarkable fact was also
noticed recently in another aggregation model namely the asymmetric random
average process with sequential updates\cite{krug,RM}, suggesting that
such unusual correlations may be more generic and less exceptional. 

In this paper we have studied the undirected CM. As mentioned in the
introduction, the directed CM also has a qualitatively similar phase
transition in the steady state though the associated critical exponents
are entirely different from the undirected one. Also, the phase boundary
in the $\rho-w$ plane of the directed CM was found be quite different from
the mean field phase boundary\cite{MKB2}. A proper understanding of the
directed model remains an outstanding challenging problem. 

We thank M. Barma, D. Dhar and S. Krishnamurthy for useful discussions. 

\newpage
\begin{figure}
\begin{center}
\leavevmode
\psfig{figure=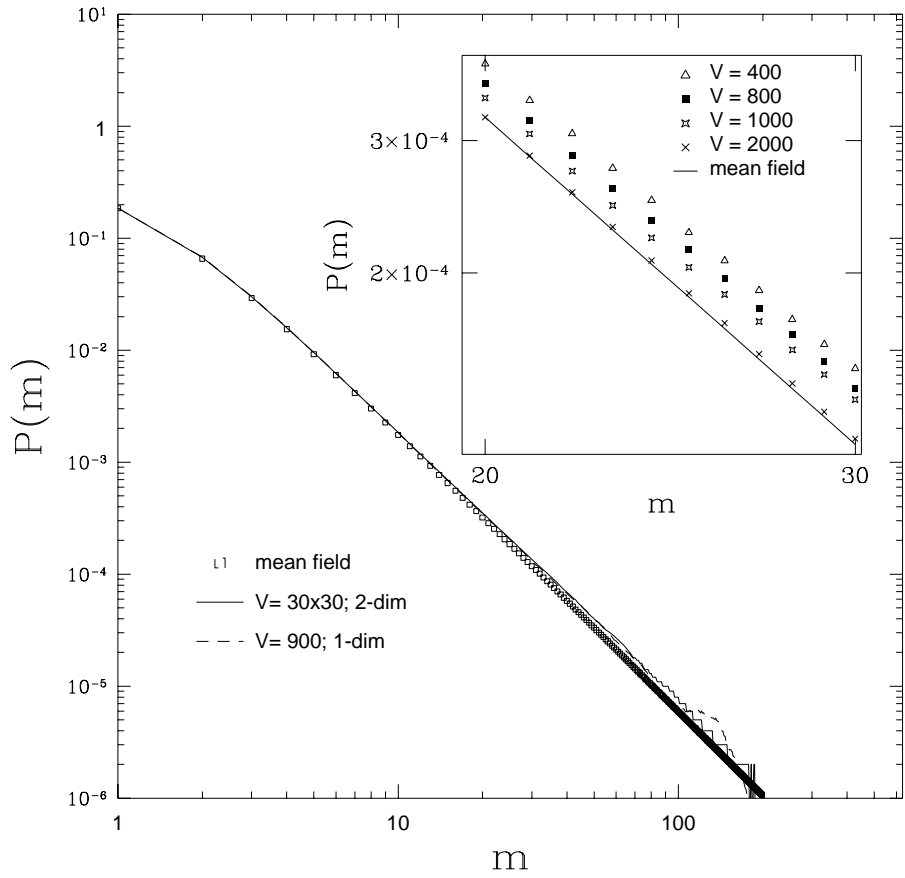,width=12cm,angle=0}
 \caption{The power law part of the steady state mass distributions $P(m)$
in $1$- and $2$-dimensions are compared with the mean field answer in the
aggregate phase $\rho>\rho_c$.  The data is for $w=3$ and $\rho=10$. The
critical value is $\rho_c=1$ for $w=3$.  In the inset panel, we show the
convergence of the probability distribution to its mean field value as the
system size is increased in one dimension.}
 \end{center}
\end{figure}

\begin{figure}
\begin{center}
\leavevmode
\psfig{figure=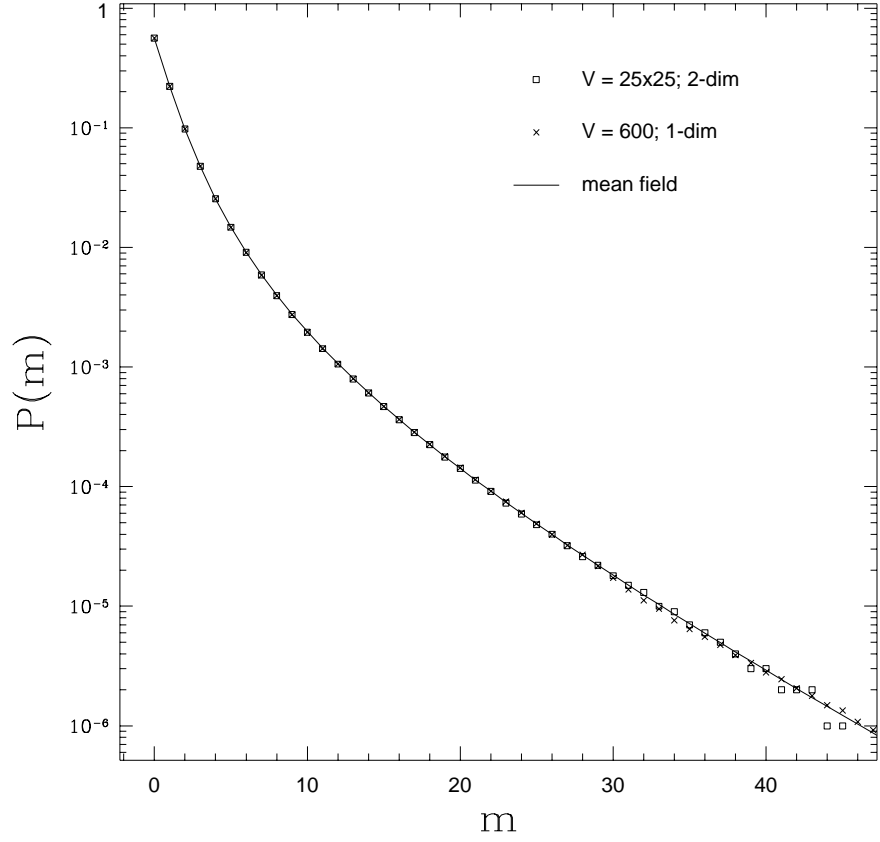,width=12cm,angle=0}
 \caption{The steady state mass distributions in $1$ and $2$ dimensions in
the exponential phase $(\rho<\rho_c)$ is compared with the mean field
answer. The data is for $w=8$ and $\rho=1$. The critical value is
$\rho_c=2$ for $w=8$.}
 \end{center}
\end{figure}

\begin{figure}
\begin{center}
\leavevmode
\psfig{figure=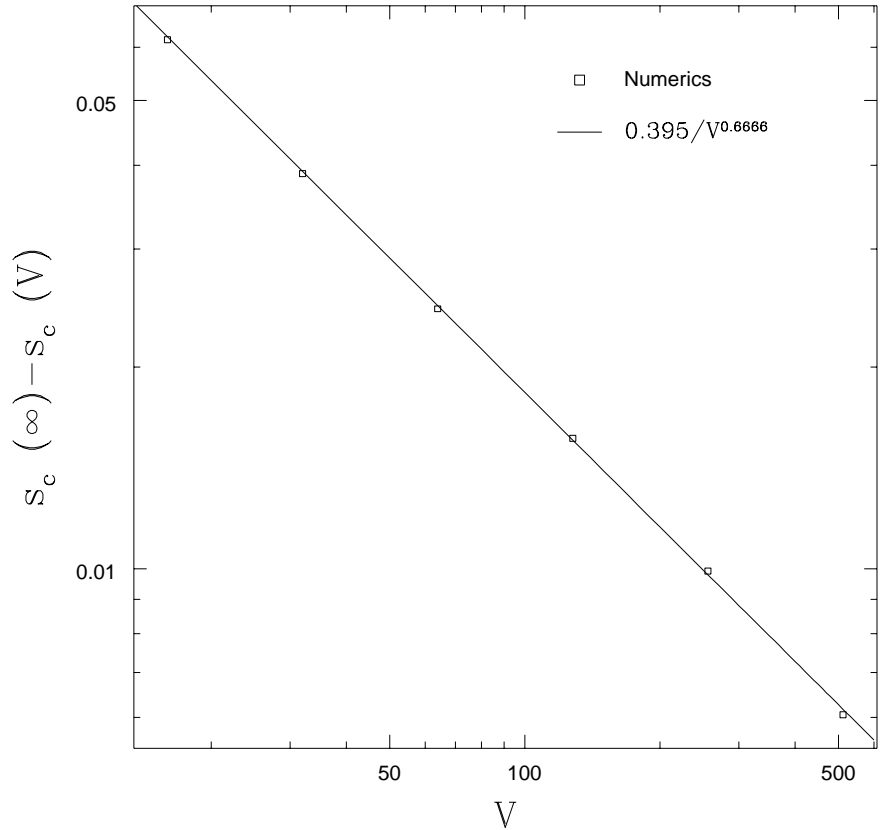,width=12cm,angle=0}
 \caption{ The probability of a site being occupied, $s_c (V)$, converges
to its asymptotic value, $s_c(\infty)$, as a power law, i. e.,
$s_c(\infty)-s_c(V)\sim 1/V^{0.66}$. The data is for lattices in
$1$-dimension}
 \end{center}
\end{figure}

\end{document}